\documentclass{PoS}
\usepackage{amssymb,amsmath}
\usepackage{upgreek}
\usepackage[permil]{overpic}
\usepackage{pict2e}
\usepackage{booktabs}
\usepackage{caption}
\usepackage{subfig}
\usepackage[capitalise]{cleveref}
\usepackage[section]{placeins}
\usepackage{cite}

\graphicspath{ {/home/mkofarag/work/Public/Presentations/Conferences/Bergen_2017/} }

\title{Angular correlation results from ALICE}

\ShortTitle{Angular correlation results from ALICE}

\author{\speaker{Monika Varga-Kofarago}\thanks{on behalf of the ALICE collaboration.}\\
        MTA Wigner RCP\\
        E-mail: \email{varga-kofarago.monika@wigner.mta.hu}}

\abstract{In heavy-ion collisions, the quark--gluon plasma is expected to be produced, which is an almost perfect liquid that made up the Universe a few microseconds after the Big Bang. In these collisions, jets are also formed from hadronizing partons with high transverse momentum, and they traverse the hot and dense medium and interact with it. Their properties can be modified by these interactions, therefore these modifications, if present, can give insight into the properties of the plasma itself. Angular correlation measurements can be used to study jets in Pb--Pb collisions in a transverse momentum ($p_{\rm T}$) regime where jets are not easily reconstructable above the fluctuating background.  Small collision systems (e.g., pp or p--Pb) can be used as reference for these measurements; however, these collisions themselves are of interest. For example, particle production mechanisms and conservation laws can be tested in these systems. Results from Pb--Pb and pp collisions recorded by the ALICE detector are presented in this paper.}

\FullConference{
12th International Workshop on High-pT Physics in the RHIC/LHC Era\\
		2-5 October, 2017\\
		University of Bergen, Bergen, Norway}

\begin{document}

\section{Introduction}
In high energy collisions, quark-antiquark pairs are produced, which fragment and hadronize into a spray of collimated hadrons, called jets. These jets interact with the hot and dense medium and their interactions can be studied in angular correlation measurements. In these measurements, the distribution of the azimuthal angle ($\Delta\varphi$) and pseudorapidity ($\Delta\eta$) difference of a trigger and an associated object is extracted. These objects can be identified or unidentified hadrons or even the jets themselves. Depending on the type of the object, different properties of the plasma and of the interactions can be studied. 

The data used for the presented measurements were collected by the ALICE detector, of which a detailed description can be found in Ref.~\cite{ALICE}. The ALICE detector has good tracking capabilities down to low transverse momentum ($p_{\rm T}$) and excellent particle identification, making it suitable for studying heavy-ion collisions as well as small collision systems (e.g., pp or p--Pb).

\section{Hadron-hadron correlations in Pb--Pb}
\begin{figure}[!bp]
  \makebox[\textwidth][c]{
    \subfloat[]{%
      \begin{overpic}[width=0.48\textwidth]{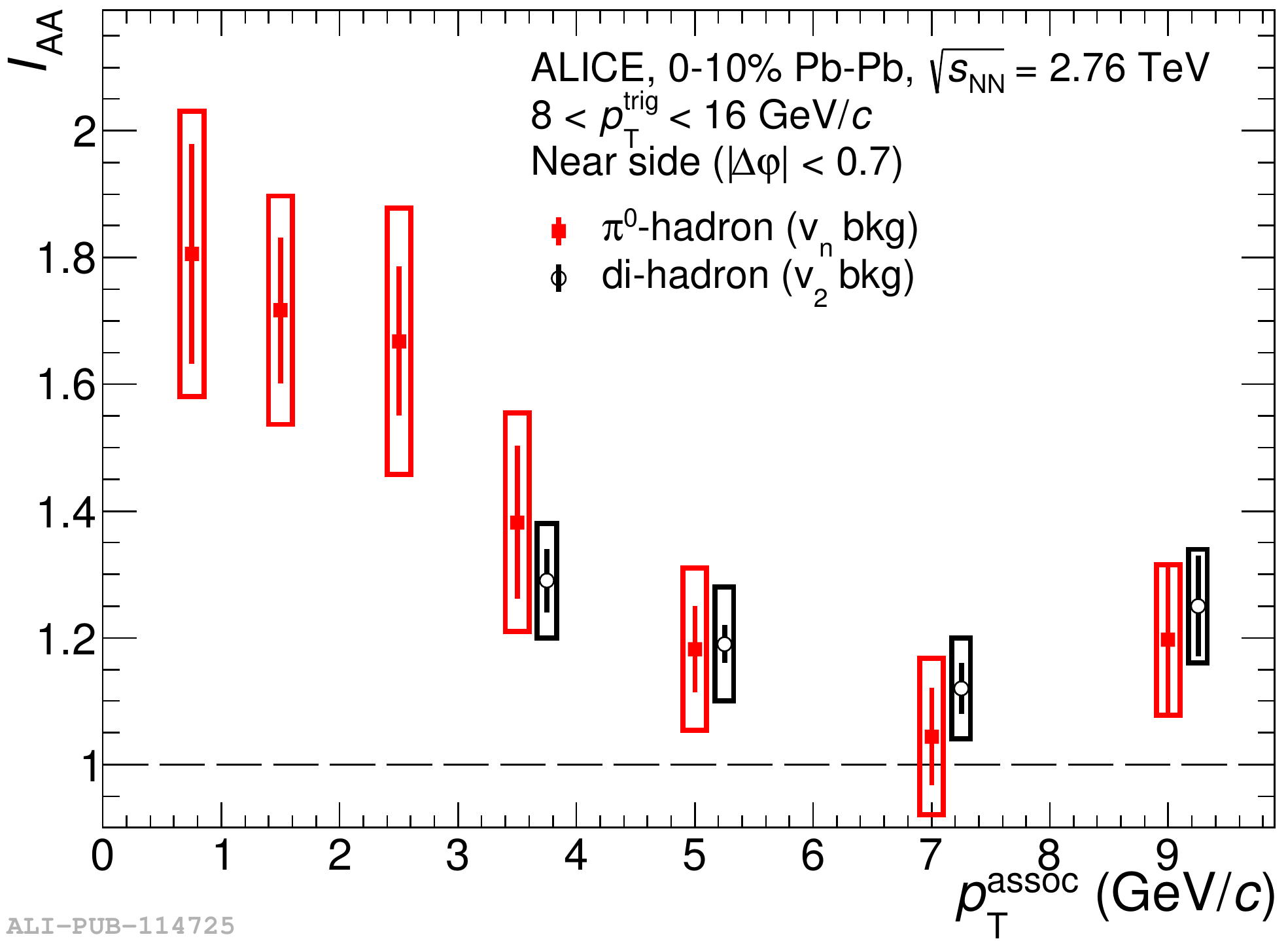}
        \label{subfig:IaaNS_0-21696}\hfill
      \end{overpic}}\hfill
    \subfloat[]{%
      \begin{overpic}[width=0.48\textwidth]{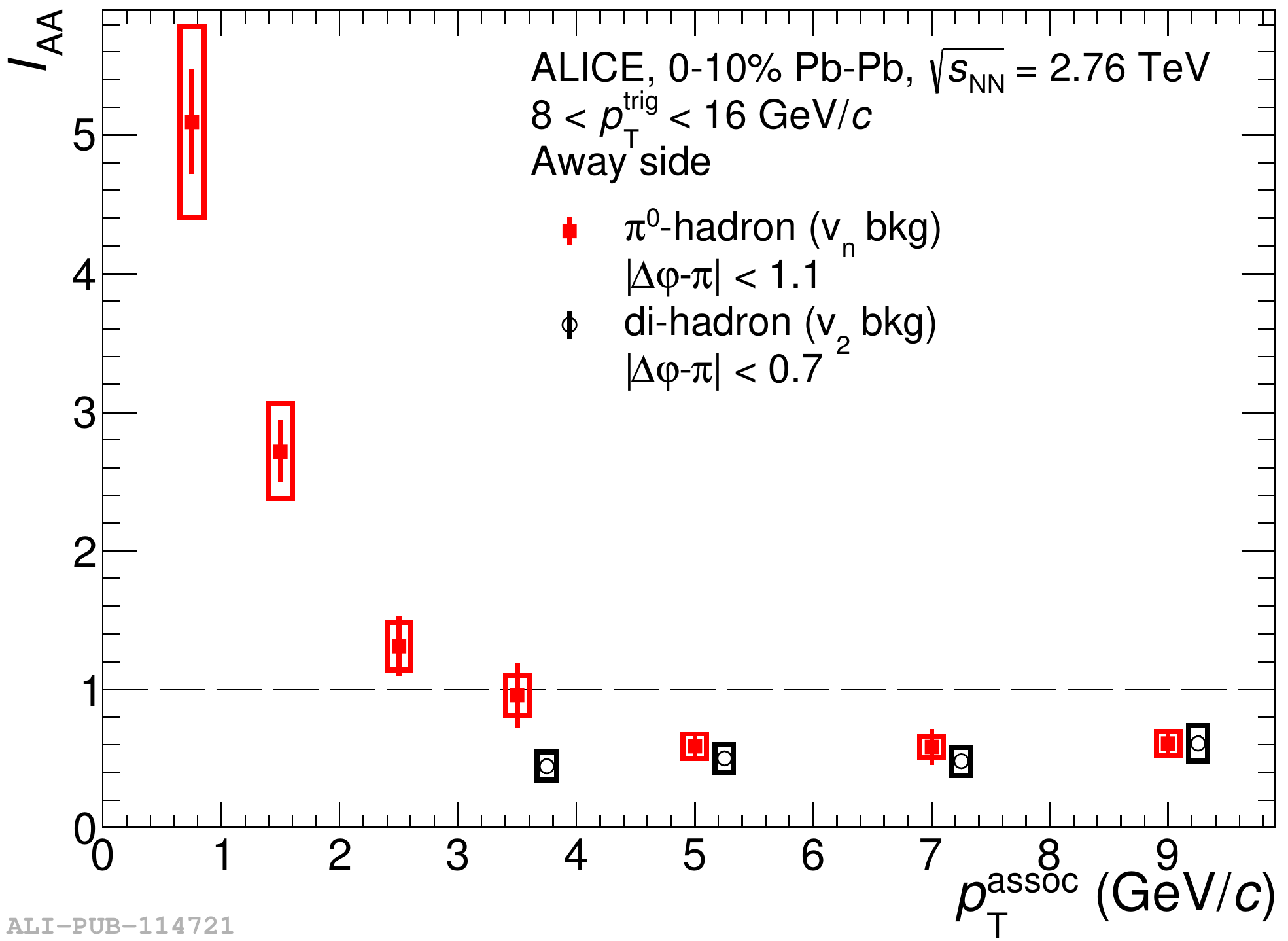}
          \label{subfig:IaaAS_0-21697}
      \end{overpic}}}
  \caption{Yield from Pb--Pb collisions divided by the yield from pp collisions (denoted as $I_{\rm AA}$) as a function of the $p_{\rm T}$ of the associated particles for hadron-hadron and $\pi^0$--hadron correlations. Panel \protect\subref{subfig:IaaNS_0-21696} shows the results from the jet peak, while panel \protect\subref{subfig:IaaAS_0-21697} shows it from the away-side structure~\cite{Adam:2016xbp}.}
  \label{fig:yieldPbPb}
\end{figure}

In angular correlation measurements of two hadrons, i.e., when the trigger and associated objects are both hadrons, jets appear as a peak around $(\Delta\varphi,\Delta\eta) = (0,0)$ (near side) and an elongated structure in $\Delta\eta$ at $\Delta\varphi=\pi$ (away side). By studying the centrality (the overlap of the two nuclei) and $p_{\rm T}$ dependence of these structures in Pb--Pb collisions, the modifications of the jets while traversing the hot and dense medium can be explored. In \cref{fig:yieldPbPb}, yields from Pb--Pb collisions of the jet peak and of the away-side structure divided by the same from pp collisions (denoted by $I_{\rm AA}$) are shown for unidentified hadron-hadron and $\pi^0$--hadron correlations as a function of the $p_{\rm T}$ of the associated particle~\cite{Adam:2016xbp}.
 A small enhancement on the near side and a small suppression on the away side can be seen at large $p_{\rm T}$, while at low $p_{\rm T}$, there is a large enhancement in both. The results are compatible in the unidentified and the identified cases in the region where both are measured. The away-side suppression can be explained by parton energy loss, while the enhancement at low $p_{\rm T}$ can originate from several sources (e.g., $k_{\rm T}$ broadening, medium-excitation or fragments from radiated gluons~\cite{Vitev,Ma,Wang,Kopeliovich,Gyulassy}). The enhancement on the near side might arise from a change of the fragmentation function or the quark-to-gluon jet ratio~\cite{Aamodt}.

\begin{figure}[!hbp]
  \begin{center}
    \begin{overpic}[width=1\textwidth]{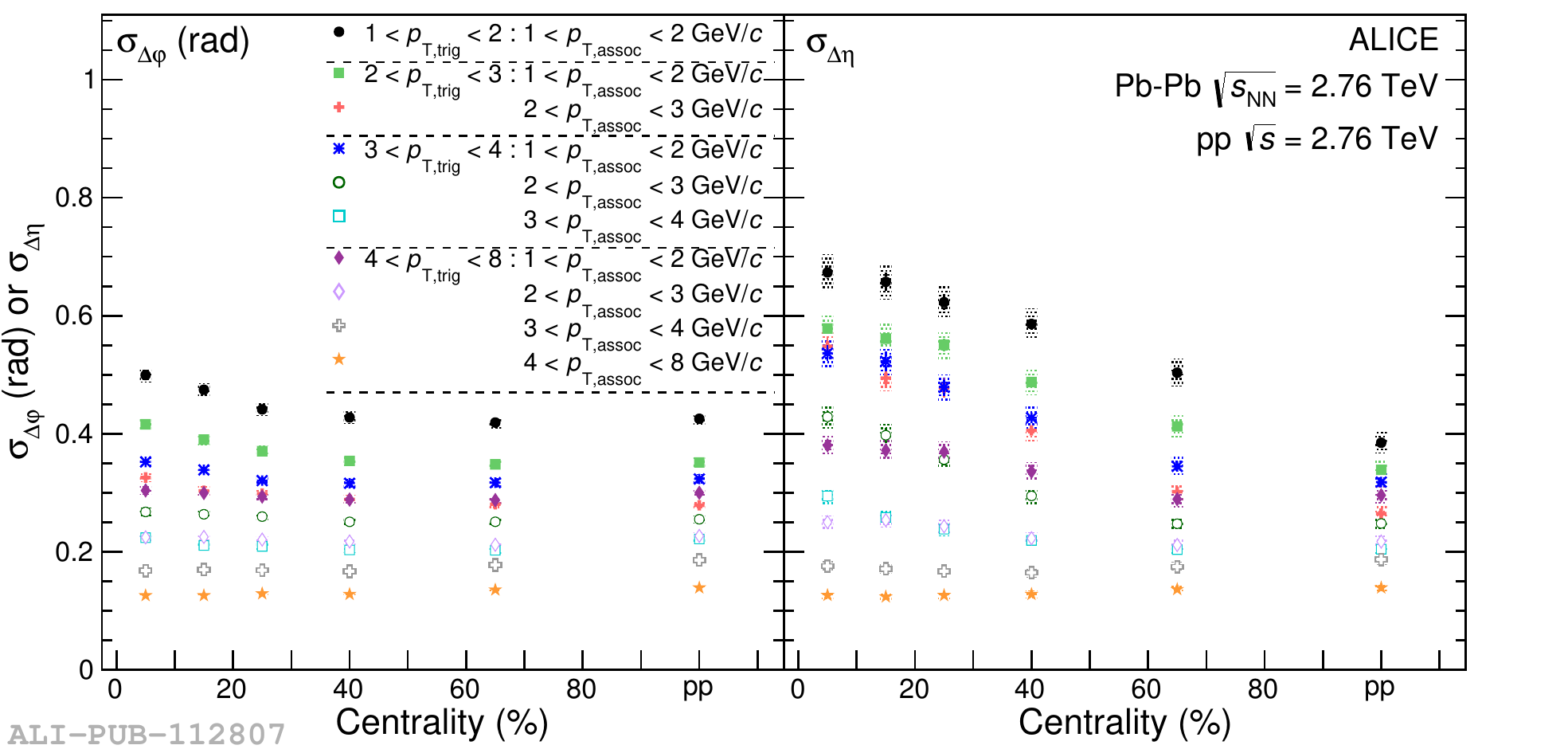}
    \end{overpic}
    \caption{Width of the jet peak from hadron-hadron collisions in the $\Delta\varphi$ (left panel) and $\Delta\eta$ (right panel) directions from Pb--Pb collisions as a function of centrality. The results from pp collisions (rightmost points) are also shown for comparison~\cite{Adam:2016tsv,Adam:2016ckp}.}
    \label{fig:widthTwoPanelPRC}
  \end{center}
\end{figure}

Apart from the yield of the jet peak, the shape of the peak can also be modified by the interaction of the partons of the jets with the medium. In \cref{fig:widthTwoPanelPRC}, the width of the peak from Pb--Pb collisions can be seen as a function of the centrality, and the results from pp collisions are shown as a reference~\cite{Adam:2016tsv,Adam:2016ckp}. The peak at low $p_{\rm T}$ gets broader towards central collisions in Pb--Pb collisions in both the $\Delta\varphi$ and the $\Delta\eta$ directions. In peripheral collisions, it converges to the value in pp collisions in the $\Delta\varphi$ direction, while in the $\Delta\eta$ direction even the peripheral Pb--Pb results are above the pp results. At low $p_{\rm T}$, the width is larger in the $\Delta\eta$ direction at the same $p_{\rm T}$ and centrality than in the $\Delta\varphi$ direction, resulting in an asymmetric peak. The multiplicity dependence of the yield and the width was also measured in pp collisions and only a very small dependence was found, therefore the use of minimum bias pp data as a reference was validated.

\begin{figure}[!htbp]
  \makebox[\textwidth][c]{
    \subfloat[]{%
      \begin{overpic}[width=0.35\textwidth]{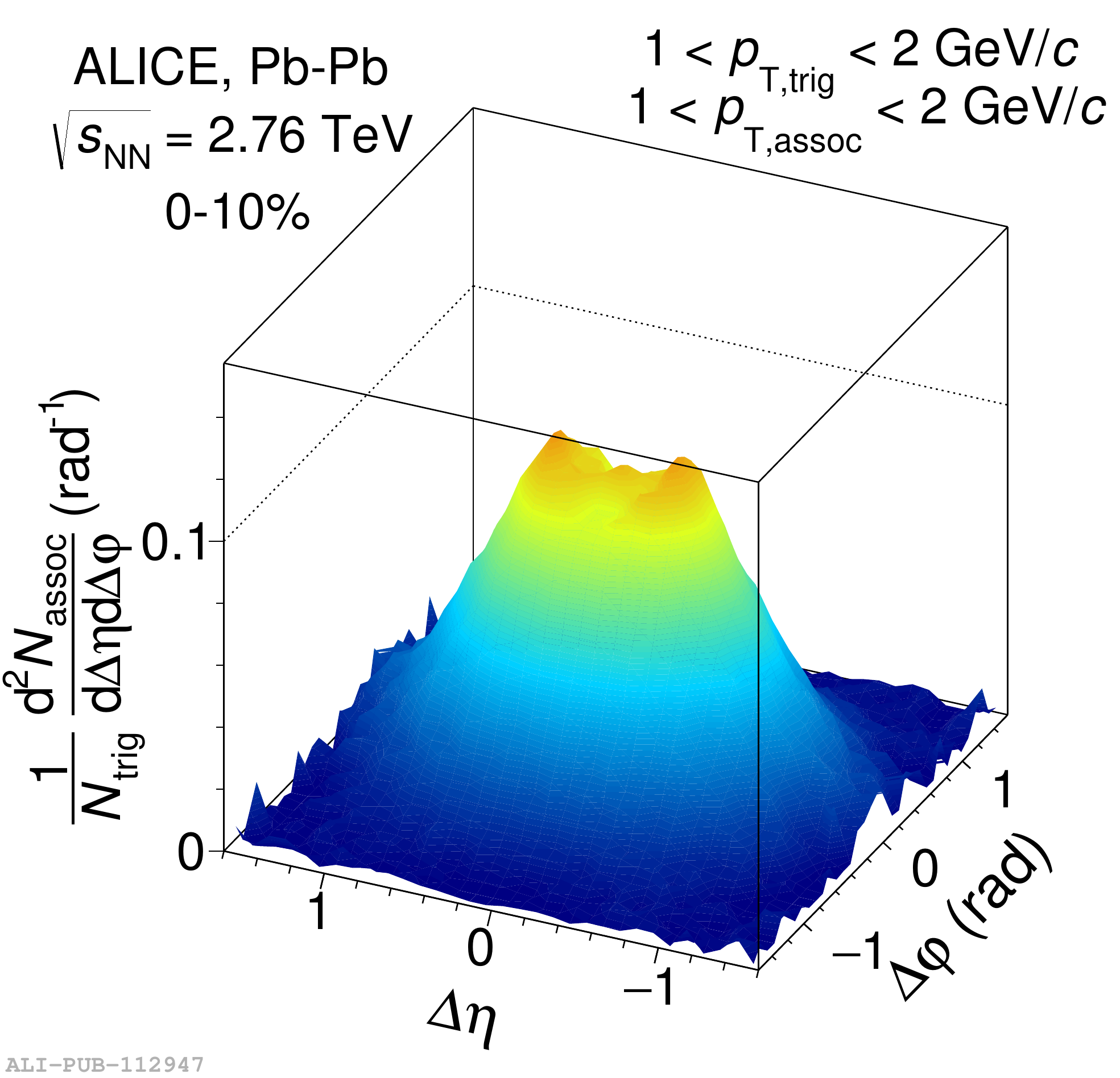}
        \label{subfig:results1c}
      \end{overpic}}\hfill
    \subfloat[]{
      \begin{overpic}[width=0.46\textwidth]{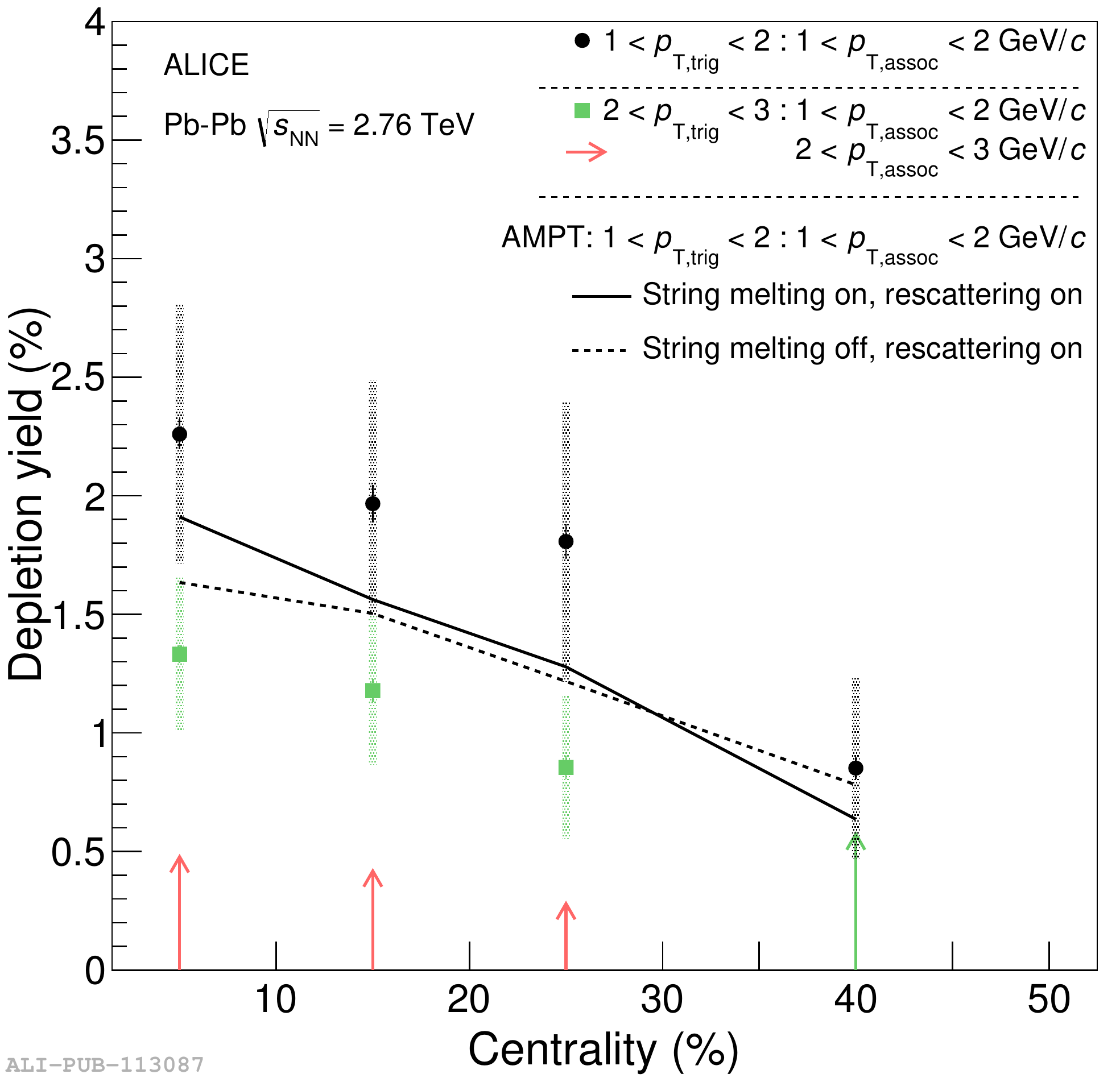}
          \label{subfig:depletion_AMPT_comparison}
      \end{overpic}}}
  \caption{Panel \protect\subref{subfig:results1c} illustrates the depletion around $(\Delta\varphi,\Delta\eta) = (0,0)$, while panel \protect\subref{subfig:depletion_AMPT_comparison} shows the missing yield in the depletion as a function of the centrality of the Pb--Pb collisions~\cite{Adam:2016tsv,Adam:2016ckp}.}
  \label{fig:depletion}
\end{figure}

In Pb--Pb collisions, at the lowest measured $p_{\rm T}$ and at the highest centrality, an unexpected depletion around $(\Delta\varphi,\Delta\eta) = (0,0)$ is seen~\cite{Adam:2016tsv,Adam:2016ckp}. In \cref{subfig:results1c}, this depletion is illustrated, while in \cref{subfig:depletion_AMPT_comparison}, the magnitude of the missing yield is presented. The results from hadron-hadron correlations in Pb--Pb collisions were compared to A Multi-Phase Transport (AMPT) model~\cite{AMPT1,AMPT2}, which exhibits both the broadening and the depletion with the settings used, and also show large radial flow. This comparison suggests therefore that these features are caused by an interplay of the jets with the flowing medium.

\section{Path-length dependence in Pb--Pb}
\begin{figure}[!htbp]
  \makebox[\textwidth][c]{
    \subfloat[]{%
      \begin{overpic}[width=0.45\textwidth]{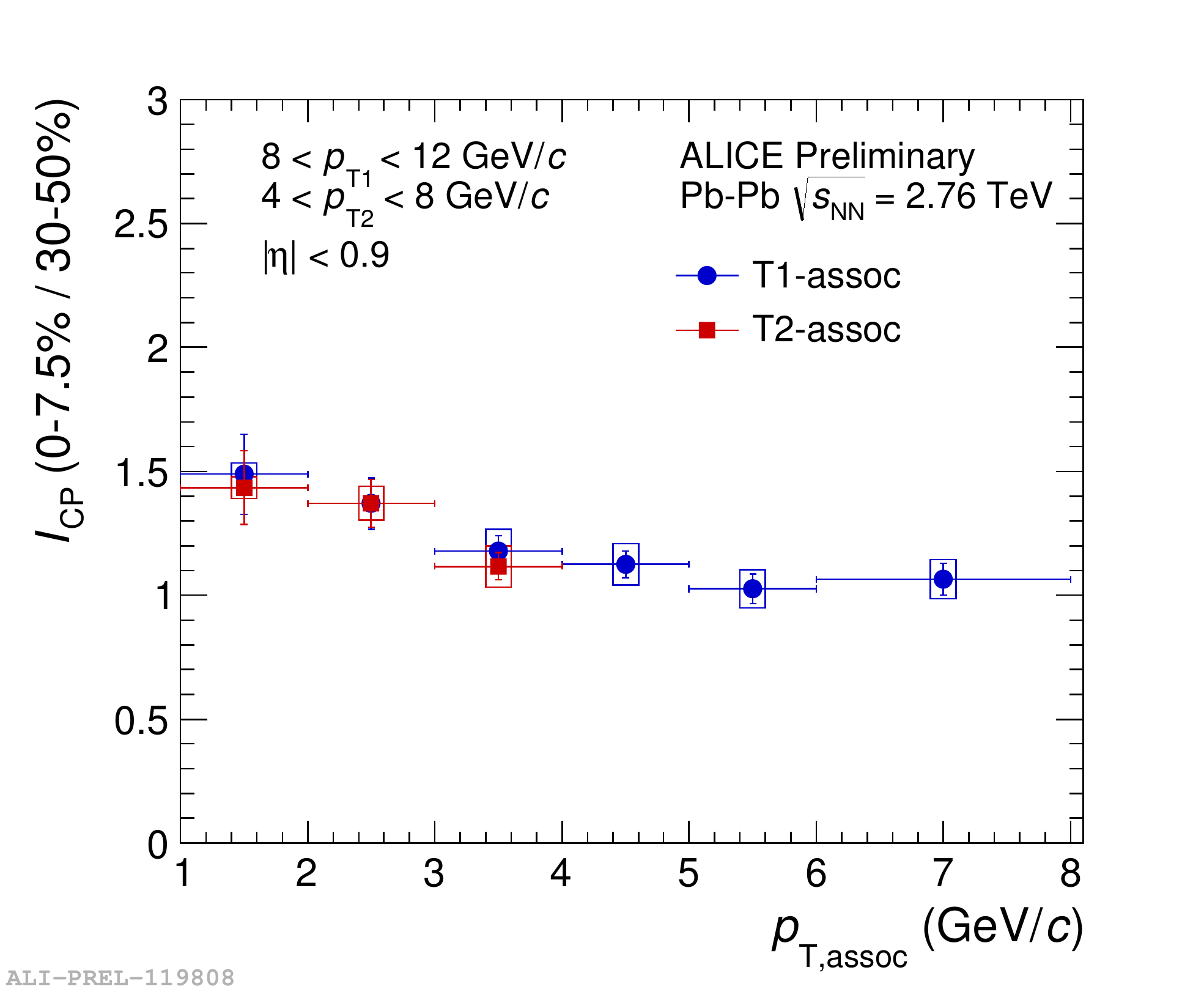}
        \label{subfig:2017-Feb-03-Icp_8_12}
      \end{overpic}}\hfill
    \subfloat[]{%
      \begin{overpic}[width=0.45\textwidth]{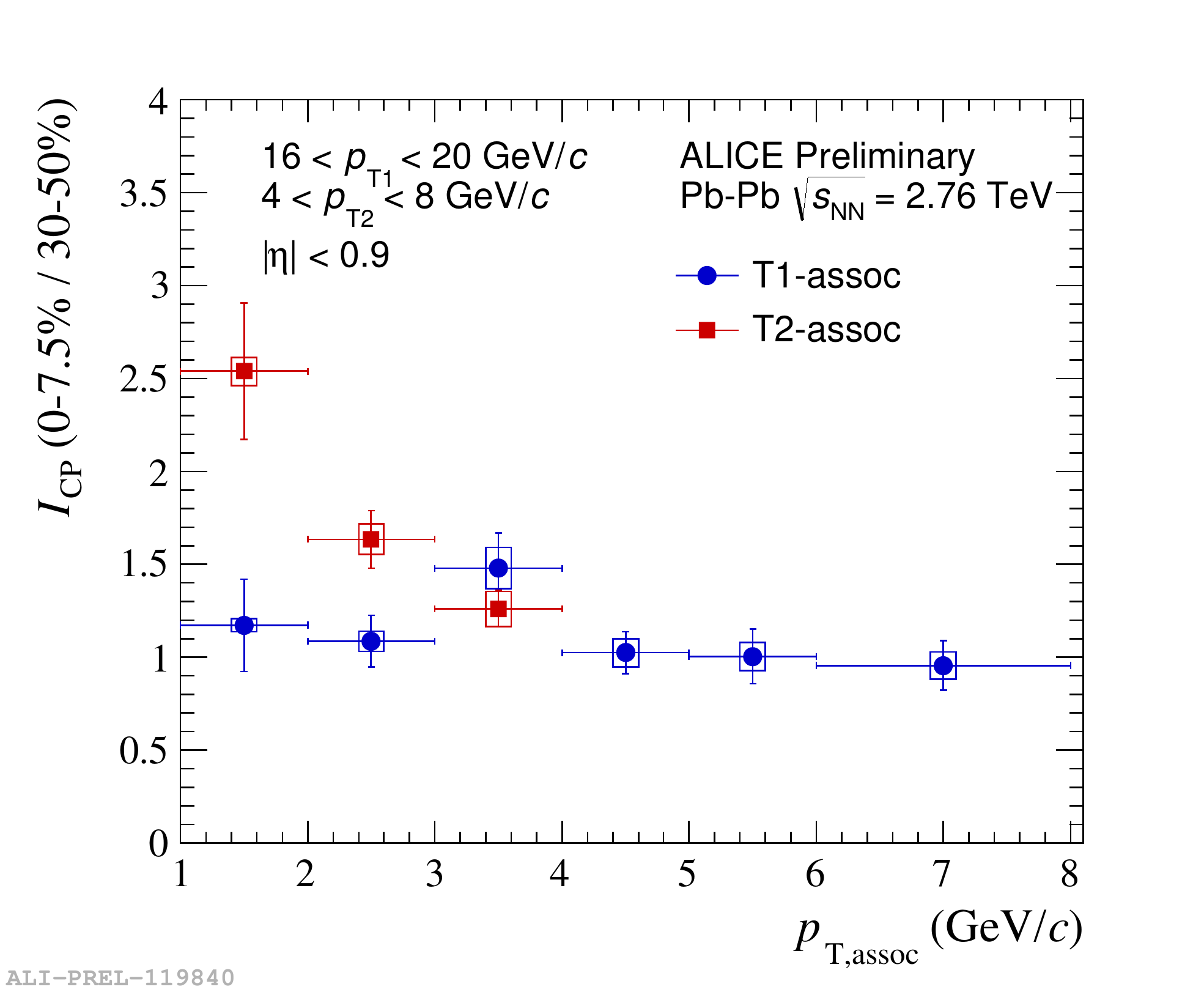}
        \label{subfig:2017-Feb-03-Icp_16_20}
      \end{overpic}}}
  \caption{Associated hadron yield from central Pb--Pb collisions divided by the same from peripheral collisions from two-plus-one correlations as a function of the $p_{\rm T}$ of the associated particles. Panel \protect\subref{subfig:2017-Feb-03-Icp_8_12} shows a case where the two trigger particles have similar $p_{\rm T}$, while panel \protect\subref{subfig:2017-Feb-03-Icp_16_20} shows an asymmetric case.}
  \label{fig:two-plus-one}
\end{figure}

The dependence on the length of the path the partons of a jet traveled in the medium can be characterized with different methods. Firstly, two-plus-one correlations are presented. In these studies, the angular distributions are calculated for two back-to-back trigger hadrons. The first trigger hadron has a higher $p_{\rm T}$, therefore it is assumed that this has traveled a shorter length in the plasma and has lost less energy than its back-to-back partner. The associated particles with this trigger particle form a peak, which corresponds to the leading jet. The associated particles with the second trigger hadron also form a peak, which corresponds to the back-to-back jet, the partons of which have traveled a longer path in the medium. The yield in central Pb--Pb collisions divided by the yield from peripheral collisions ($I_{\rm CP}$) from the peak for both trigger particles is shown in \cref{fig:two-plus-one}. In the case when the two trigger particles have similar $p_{\rm T}$ (\cref{subfig:2017-Feb-03-Icp_8_12}), the $I_{\rm CP}$ corresponding to the two trigger particles are the same, while in the case when the second trigger particle has a much lower $p_{\rm T}$ than the first (\cref{subfig:2017-Feb-03-Icp_16_20}), the $I_{\rm CP}$ belonging to the second trigger particle is enhanced at low~$p_{\rm T}$. This can be explained by a path length dependent suppression, but other explanations (e.g., a bias on the trigger selection) are also possible.

\begin{figure}[!htbp]
  \makebox[\textwidth][c]{
    \subfloat[]{%
      \begin{overpic}[width=0.43\textwidth]{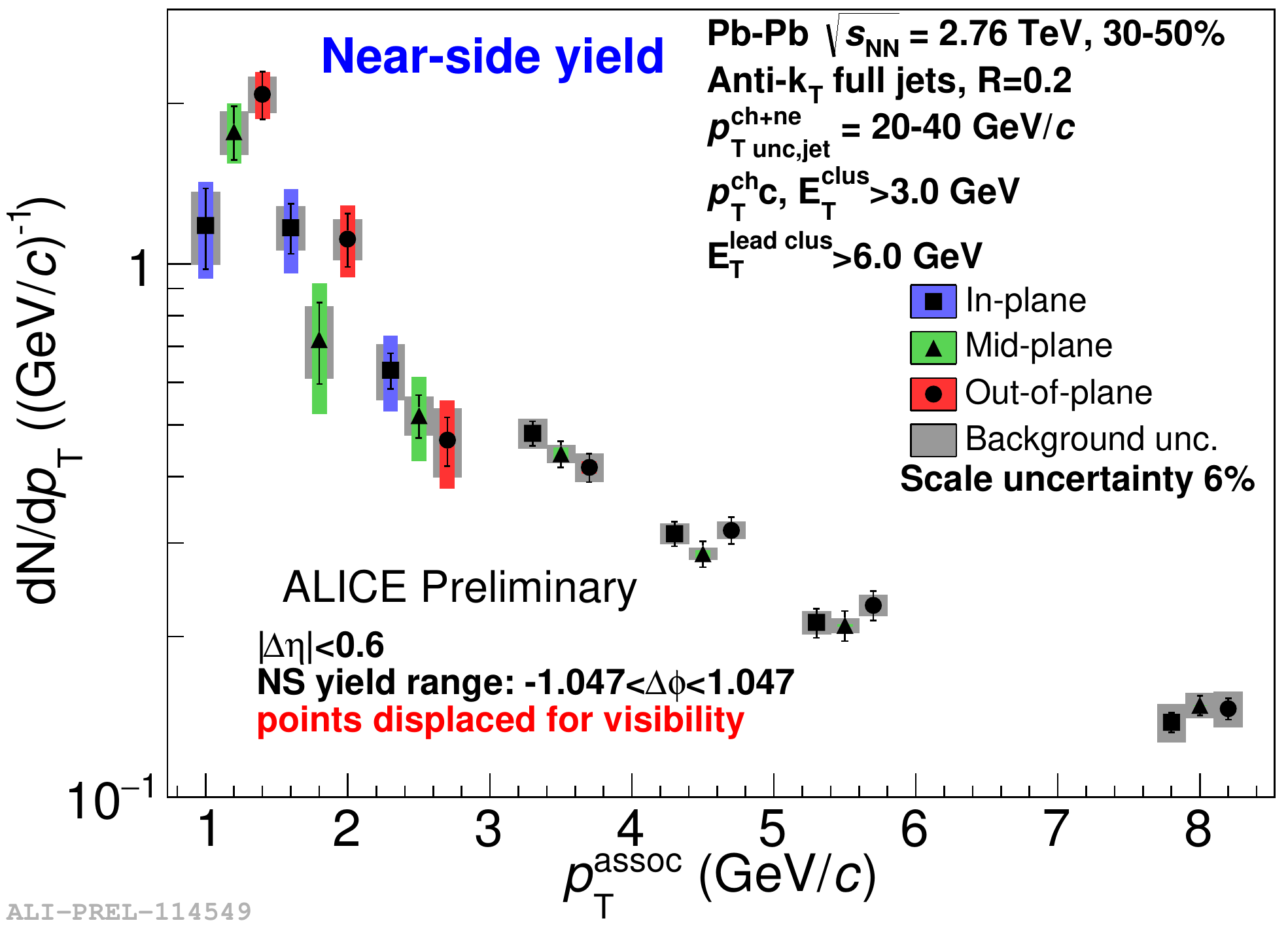}
        \label{subfig:NearSideYield}
      \end{overpic}}\hfill
    \subfloat[]{%
      \begin{overpic}[width=0.43\textwidth]{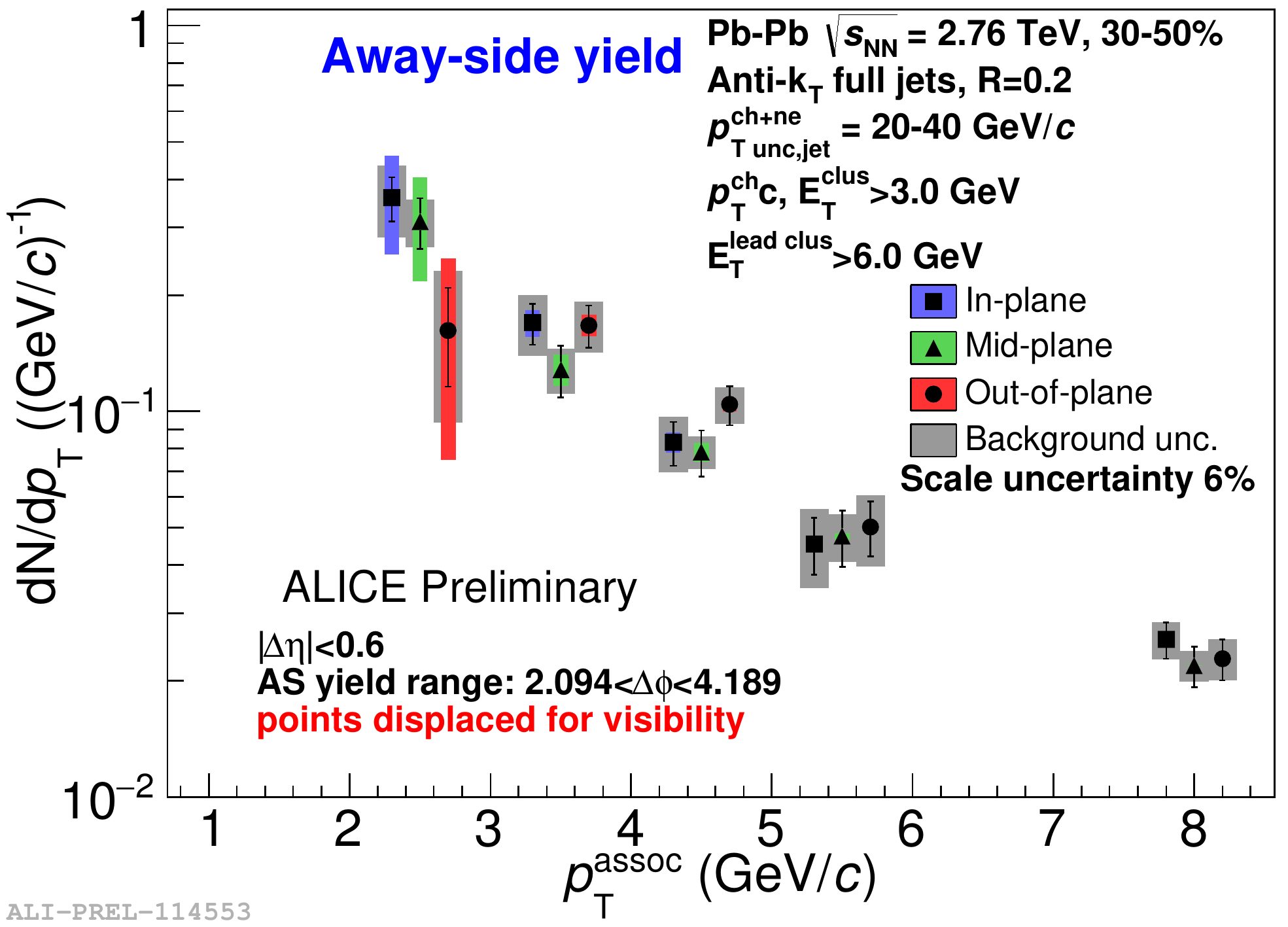}
        \label{subfig:AwaySideYield}
      \end{overpic}}}
  \caption{Associated hadron yield for a jet trigger in Pb--Pb collisions where the direction of the jet is restricted with respect to the event plane. In panel \protect\subref{subfig:NearSideYield}, the yield on the near side is shown, while panel \protect\subref{subfig:AwaySideYield} presents the away side.}
  \label{fig:jet-hadron}
\end{figure}

Secondly, the path length dependence of jets can be studied by taking jets as the trigger objects and restricting their direction with respect to the direction of the event plane (the plane defined by the direction of the colliding nuclei and the vector connecting the centers of the colliding nuclei). In this way, jets that are parallel with the event plane travel a smaller path in the medium than jets that go perpendicular to the event plane. In \cref{fig:jet-hadron}, the yield of hadrons associated with jets is presented for three categories of jets depending on their angle with respect to the event plane. The yields agree within uncertainties for the three cases, therefore no dependence on the path length is found. This suggests that the modifications seen from the two-plus-one correlations are connected to another effect.

\section{Small systems}
\begin{figure}[!htbp]
  \makebox[\textwidth][c]{
    \subfloat[]{%
      \begin{overpic}[width=0.44\textwidth]{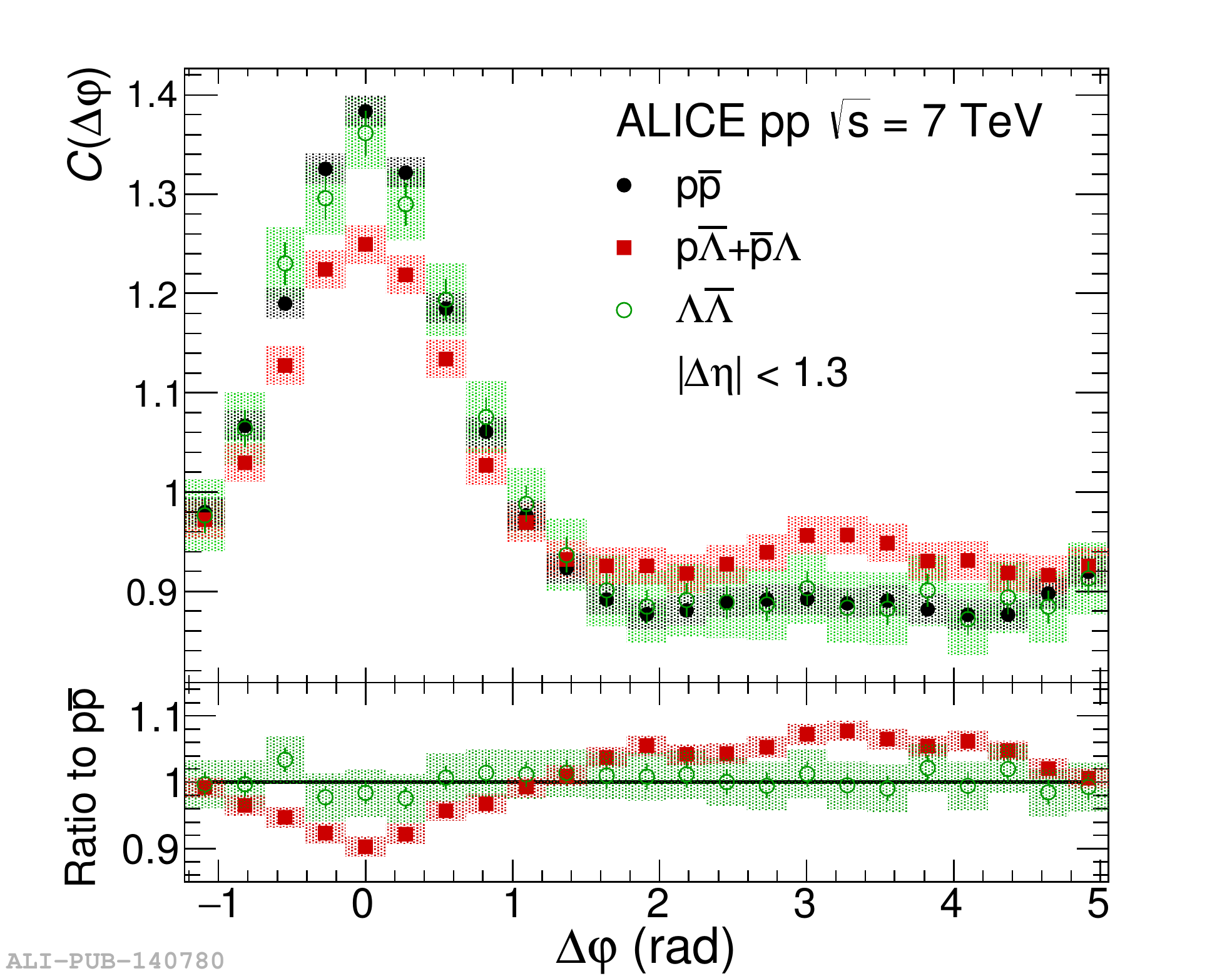}
        \label{subfig:BaryonsToPP_UnLikeSign}
      \end{overpic}}\hfill
    \subfloat[]{%
      \begin{overpic}[width=0.44\textwidth]{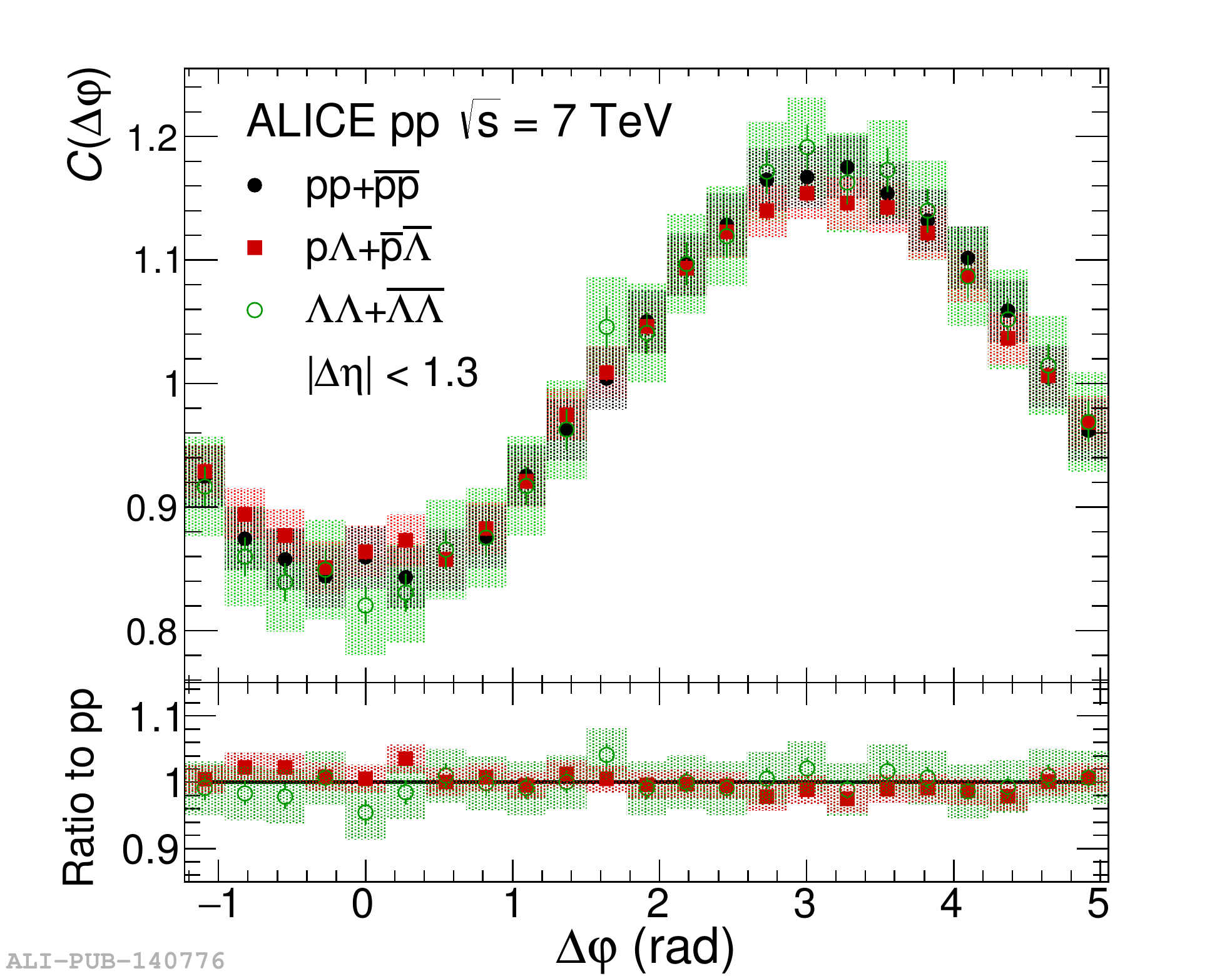}
          \label{subfig:BaryonsToPP_LikeSign}
      \end{overpic}}}
  \caption{Projection of the associated yield for identified baryon-baryon correlations. In panel \protect\subref{subfig:BaryonsToPP_UnLikeSign} unlike-sign baryons are considered, while in panel \protect\subref{subfig:BaryonsToPP_LikeSign}, the like-sign case is shown~\cite{Adam:2016iwf}.}
  \label{fig:pp_identified}
\end{figure}

Angular correlations of identified particles in small systems can be used to study the production mechanisms and different conservation laws of the studied particles. In \cref{fig:pp_identified}, identified hadron-hadron correlations are presented from pp collisions~\cite{Adam:2016iwf}. When mesons (e.g., pions or kaons) are considered, a jet peak is visible, as expected, in both the like-sign and the unlike-sign case (not shown here). However, when baryons are considered, the jet peak is only visible in the unlike-sign case (\cref{subfig:BaryonsToPP_UnLikeSign}). In the like-sign case, a depletion is present instead of the peak (\cref{subfig:BaryonsToPP_LikeSign}). The size of this depletion is the same if protons or $\Lambda$ particles are considered, and it is not described by models, therefore its origin is currently not understood.

\section{Summary}
Angular correlation measurements are powerful tools to study the interaction of the partons of jets with the medium in heavy-ion correlations, and to study the production mechanisms and conservation laws in small systems. In Pb--Pb collisions, an enhancement of the jet-peak yield and a broadening of the jet peak at low transverse momentum in central collisions were presented together with an unexpected depletion around $(\Delta\varphi,\Delta\eta) = (0,0)$. Furthermore, studies on the path length dependence of the measured quantities were shown, with no clear path length dependence observed. Finally, measurements from pp collisions were presented, where an unexpected depletion of the jet peak is visible when baryon-baryon like-sign correlations are studied.

\FloatBarrier

\end{document}